\documentclass[onecolumn,aps,letterpaper]{revtex4}
\usepackage{epsfig}

\newcommand{\BABARPubYear}    {09}

\newcommand{\BABARConfNumber} {001}
\newcommand{\SLACPubNumber} {13543}

\input babarsym

\def\cpoddhiggs      {\ensuremath{{A^{0}}}\xspace}

\def\n1Spipi     {\ensuremath{}\xspace}

\def\beq{\begin{equation}}
\def\eeq{\end{equation}}
\def\bea{\begin{eqnarray}}
\def\eea{\end{eqnarray}}
\def\bq{\begin{quote}}
\def\eq{\end{quote}}
\def\bi{\begin{itemize}}
\def\ei{\end{itemize}}
\def\bc{\begin{center}}
\def\ec{\end{center}}

\newcommand{\eg}{{\em e.g.}}

\def\etal{{\em et al.}}

\newcommand{\higgsmass}{\ensuremath{m_{A^0}}\xspace}

\newcommand{\redmass}{\ensuremath{m_R}\xspace}

\long\def\inst#1{\par\nobreak\kern 4pt\nobreak
    {\it #1}\par\vskip 10pt plus 3pt minus 3pt}

\begin{document}

\thispagestyle{empty}

\begin{flushright}
\babar-CONF-\BABARPubYear/\BABARConfNumber \\
SLAC-PUB-\SLACPubNumber \\
February 2009 \\
\end{flushright}

\par\vskip 5cm

\begin{center}
\Large \bf 
Search for Dimuon Decays of a Light Scalar in Radiative
Transitions $\Upsilon(3S)\to\gamma\cpoddhiggs$
\end{center}
\bigskip

\begin{center}
\large The \babar\ Collaboration\\
\mbox{ }\\
February 12, 2009
\end{center}
\bigskip \bigskip

\begin{center}
\large \bf Abstract
\end{center}
The fundamental nature of mass is one of the greatest mysteries in
physics. The Higgs mechanism is a theoretically appealing way to
account for the different masses of elementary particles and implies
the existence of a new, yet unseen particle, the Higgs boson. We
search for evidence of a light scalar (\eg\ a Higgs boson) in the
radiative decays of 
the narrow \Y3S resonance: $\Y3S\to\gamma\cpoddhiggs$,
$\cpoddhiggs\to\mu^+\mu^-$. Such an object 
appears in extensions of the Standard
Model, where a light \CP-odd Higgs boson naturally couples strongly to
$b$-quarks. 
We find no evidence for such processes in 
a sample of $122\times10^6$ $\Upsilon(3S)$  decays collected
by the \babar\ collaboration at the \pep2\ B-factory, and set 90\%
C.L. upper limits 
on the branching fraction product
$\mathcal{B}(\Upsilon(3S)\to\gamma A^0)\times\mathcal{B}(A^0\to\mu^+\mu^-)$ at
$(0.25-5.2)\times10^{-6}$ in the mass range $0.212\le m_{A^0}\le9.3$\,GeV. 
We also set a limit on the dimuon branching fraction of the $\eta_b$
meson $\BR(\eta_b\to\mu^+\mu^-)<0.8\%$ at 90\% C.L.
The results are preliminary. 

\vfill
\begin{center}
Submitted to Aspen 2009 Winter Conference,
February 8---14,  2009, Aspen, CO
\end{center}

\vspace{1.0cm}
\begin{center}
{\em SLAC National Accelerator Laboratory, Stanford University, 
Stanford, CA 94309} \\ \vspace{0.1cm}\hrule\vspace{0.1cm}
Work supported in part by Department of Energy contract DE-AC02-76SF00515.
\end{center}

\newpage

%
%
%
\author{B.~Aubert}
\author{Y.~Karyotakis}
\author{J.~P.~Lees}
\author{V.~Poireau}
\author{E.~Prencipe}
\author{X.~Prudent}
\author{V.~Tisserand}
\affiliation{Laboratoire d'Annecy-le-Vieux de Physique des Particules (LAPP), Universit\'e de Savoie, CNRS/IN2P3, F-74941 Annecy-Le-Vieux, France }
\author{J.~Garra~Tico}
\author{E.~Grauges}
\affiliation{Universitat de Barcelona, Facultat de Fisica, Departament ECM, E-08028 Barcelona, Spain }
\author{M.~Martinelli}
\author{A.~Palano$^{ab}$ }
\author{M.~Pappagallo$^{ab}$ }
\affiliation{INFN Sezione di Bari$^{a}$; Dipartmento di Fisica, Universit\`a di Bari$^{b}$, I-70126 Bari, Italy }
% ...
\collaboration{The \babar\ Collaboration}
\noaffiliation

\maketitle
\newpage

\section{INTRODUCTION}
\label{sec:Introduction}

The concept of mass is one of the most intuitive ideas in physics since it 
is present in everyday human experience.  Yet the fundamental nature of mass
remains one of the greatest mysteries in physics.  The Higgs mechanism is a
theoretically appealing way to account for the different masses of elementary 
particles~\cite{ref:Higgs}. 
The Higgs mechanism implies the existence of
at least one new particle 
called the Higgs boson, which is the only Standard Model
(SM)~\cite{ref:SM} particle yet to be observed.  If it is found, its
discovery will have a profound effect  
on our fundamental understanding of matter.
A single Standard Model Higgs
boson is required to be heavy, with the mass constrained by 
direct searches to $m_{H} >114.4$\,GeV \cite{Barate:2003sz}
and $m_{H}\ne 170$\,GeV \cite{Herndon:ICHEP08},
and by
precision electroweak measurements to 
$m_{H} = 129^{+74}_{-49}$\,GeV \cite{LEPSLC:2005ema}.

The Standard Model and the simplest electroweak symmetry breaking
scenario suffer from quadratic divergences in the radiative
corrections to the mass parameter of the Higgs potential. 
Several theories beyond the Standard Model that regulate these
divergences have been proposed. 
Supersymmetry~\cite{ref:SUSY} is one such model; however, in its simplest form 
(the Minimal Supersymmetric Standard Model, MSSM) questions of
parameter fine-tuning and ``naturalness'' of the Higgs mass scale
remain. 

Theoretical efforts to solve unattractive features of MSSM often
result in models that introduce additional Higgs fields, with one of
them naturally light. For instance, the Next-to-Minimal Supersymmetric
Standard Model (NMSSM)~\cite{Dermisek:2005ar} introduces a singlet
Higgs field. A linear combination of this singlet state with a member
of the electroweak doublet produces a \CP-odd Higgs state $A^0$ 
whose mass is not required to be large. 
Direct searches typically constrain \higgsmass\ to be
below $2m_b$~\cite{Dermisek:2006} making it accessible to decays of
$\Upsilon$ resonances. 
An ideal place to search for such \CP-odd Higgs would be
$\Upsilon \to \gamma \cpoddhiggs$, as originally proposed by Wilczek
\cite{Wilczek:1977zn}.  A study of the NMSSM parameter
space \cite{Dermisek:2006py} predicts the
branching fraction to this final state to be as high as $10^{-4}$. 

Other new physics models,
motivated by astrophysical observations, predict similar light
states. One recent example~\cite{NomuraThaler} proposes a light
axion-like pseudoscalar boson $a$ decaying predominantly to leptons and
predicts the branching fraction $\BR(\Upsilon\rightarrow \gamma\ a)$ to
be between $10^{-6}$\textendash $10^{-5}$~\cite{NomuraThaler}.
Empirical
motivation for a light Higgs search comes from the HyperCP
experiment~\cite{HyperCP}.  HyperCP observed three anomalous events
in the $\Sigma\to p\mu^+\mu^-$ final state, that have been 
interpreted as a light scalar with mass of $214.3$\,\mev decaying
into a pair of muons~\cite{XJG}.
The large datasets available at \babar\ allow us to place stringent
constraints on such models.

If a light scalar \cpoddhiggs\ exists, the pattern of its decays would depend
on its mass. Assuming no invisible (neutralino) decays~\cite{BAD2073},
for low masses 
$\higgsmass<2 m_\tau$, relevant for the HyperCP~\cite{HyperCP} and
axion~\cite{NomuraThaler} interpretations, the dominant decay mode
should be $\cpoddhiggs\to\mu^+\mu^-$. Significantly above the tau threshold,
$\cpoddhiggs\to\tau^+\tau^-$ would dominate, and the hadronic decays may also
be significant. This analysis searches for the radiative production of
\cpoddhiggs\ in \Y3S decays, with \cpoddhiggs\ decaying into muons: 
\begin{displaymath}
\Y3S\to\gamma \cpoddhiggs;\ \cpoddhiggs\to\mu^+\mu^-
\end{displaymath}

The current best limit on the branching fraction 
$\mathcal{B}(\Upsilon\to\gamma\cpoddhiggs)$ with
$\cpoddhiggs\to\mu^+\mu^-$ comes from a measurement by the CLEO
collaboration on
$\Upsilon(1S)$~\cite{CLEO2008}. The quoted limits 
on $\BR(\Y1S\to\gamma\cpoddhiggs)\times\BR(\cpoddhiggs\to\mu^+\mu^-)$ 
are in the range 
(1-20)$\times10^{-6}$ for $\higgsmass<3.6$\,\gev. There are currently
no competitive 
measurements at the higher-mass $\Upsilon$ resonances or for the
values of \higgsmass\ above the $\tau\tau$ threshold. 

In the following, we describe a search for a resonance in the
dimuon invariant mass distribution for fully reconstructed final state
$\Y3S\to\gamma(\mu^+\mu^-)$. We assume that the decay width of
the resonance is negligibly 
small compared to experimental resolution, as
expected~\cite{NomuraThaler,ref:Lozano} for \higgsmass\ sufficiently
far from the mass of the $\eta_b$~\cite{ref:etab}. 
We also assume that the resonance is a scalar (or pseudo-scalar)
particle; while significance of any peak does not depend on this
assumption, the signal efficiency and, therefore, the extracted
branching fractions are computed for a spin-0 particle. 
In addition, following
the recent discovery of the $\eta_b$ meson in \Y3S
decays~\cite{ref:etab}, we look for the leptonic decay of the $\eta_b$
through the chain $\Y3S\to\gamma\eta_b$,
$\eta_b\to\mu^+\mu^-$. If the recently discovered state is the
conventional quark-antiquark $\eta_b$ meson, its leptonic width is
expected to be negligible. Thus, setting a limit on the dimuon
branching fraction sheds some light on the nature of the recently
discovered state. We assume $\Gamma(\eta_b)=10$\,MeV, which is
expected in most theoretical models and is 
consistent with \babar\ results~\cite{ref:etab}.

\section{THE \babar\ DETECTOR AND DATASET}
\label{sec:babar}

We search for
two-body transitions $\Upsilon(3S)\to\gamma\cpoddhiggs$, followed by 
decay $\cpoddhiggs\to\mu^+\mu^-$ 
 in a sample of $(121.8\pm 1.2)\times10^6$
$\Upsilon(3S)$ 
decays collected with the \babar\ detector
at the \pep2\ asymmetric-energy \epem\ collider at the Stanford Linear
Accelerator Center. The data were
collected at the nominal center-of-mass (CM) energy $E_\mathrm{cm}=10.355$\,GeV.
The CM frame was boosted relative to the
detector approximately along the detector's magnetic field axis by
$\beta_z=0.469$. 

We use a sample of $78.5\,\mathrm{fb}^{-1}$
accumulated on \Y4S resonance (\Y4S sample) for studies of the continuum
backgrounds; since \Y4S is three orders of magnitude broader than
\Y3S, the branching fraction $\Y4S\to\gamma\cpoddhiggs$ is expected to
be negligible. 
For characterization of the background events and selection
optimization we also use a sample of
$2.4\,\mathrm{fb}^{-1}$ collected $30$\,MeV below the \Y3S
resonance.

Since the \babar\ detector is described in detail
elsewhere~\cite{detector}, 
only the components of the detector crucial to this analysis are
 summarized below. 
Charged particle tracking is provided by a five-layer double-sided silicon
vertex tracker (SVT) and a 40-layer drift chamber (DCH). 
Photons and neutral pions are identified and measured using
the electromagnetic calorimeter (EMC), which comprises 6580 thallium-doped CsI
crystals. These systems are mounted inside a 1.5-T solenoidal
superconducting magnet. 
The Instrumented Flux Return (IFR) forms the return yoke of
the superconducting coil, instrumented in the central barrel region with
limited streamer tubes, and in the endcap regions with the
resistive-plate chambers, 
 for the identification 
of muons and the detection of clusters produced
by neutral hadrons. 
We use the GEANT~\cite{geant} software to simulate interactions of particles
traversing the \babar\ detector, taking into account the varying
detector conditions and beam backgrounds. 

\section{EVENT SELECTION}
\label{sec:Analysis}

We select events with exactly two oppositely-charged tracks and a
single energetic photon with a CM energy
$E^{*}_\gamma\ge0.5$\,\gev. We allow other photons to be present in the
event as long as their CM energies are below $0.5$\,\gev. We assign a
muon mass hypothesis to the two tracks (henceforth referred to as muon
candidates), and require
that they form a geometric vertex with the $\chi^2_\mathrm{vtx}<20$
(for 1 degree of freedom),
displaced transversely by 
at most $2$\,cm from the nominal location of the $e^+e^-$ interaction
region. We perform 
a kinematic fit to the \Y3S candidate formed from the two muon
candidates and the energetic photon, constraining the CM energy of the \Y3S
candidate, within the beam energy spread, to the total
beam energy $\sqrt{s}$. We also 
assume that the \Y3S candidate originates from the interaction
region. The kinematic 
fit improves the invariant mass resolution of the muon pair. We
place a requirement on the kinematic fit $\chi^2_{\Y3S}<39$ (for 6
degrees of freedom), which
corresponds to the probability to reject good kinematic fits of less
than $10^{-6}$. The kinematic fit $\chi^2$, together with a
requirement that the total mass of the $\Y3S$ candidate is within
2\,\gev of $\sqrt{s}$, suppresses background events with more than two
muons and a photon in the final state, such as cascade decays 
$\Y3S\to\gamma\chi_b(2P)\to\gamma\gamma\Y1S\to\gamma\gamma\mu^+\mu^-$
etc. We further  
require that the momentum of the dimuon candidate
$\cpoddhiggs$ and the photon direction are back-to-back in the CM frame to
within $0.07$ radians, and select events in which the cosine of the angle
between the muon direction and $\cpoddhiggs$ direction in the center
of mass of $\cpoddhiggs$ is less than $0.88$. We reject events in
which neither muon candidate is positively identified in the IFR. 

The kinematic selection described above is highly efficient for signal
events. After the selection, the backgrounds are dominated by two
types of QED processes: ``continuum'' $e^+e^-\to\gamma\mu^+\mu^-$
events in which a photon is emitted in the initial or final state, and
the initial-state radiation (ISR) production of the vector mesons
$J/\psi$, $\psi(2S)$, and \Y1S, which subsequently decay into muon
pairs. In order to suppress contributions from ISR-produced
$\rho^0\to\pi^+\pi^-$ and $\phi\to K^+K^-$ final states in which a
pion or a kaon is misidentified as a muon or decays (\eg\ through $K^+\to\mu^+\nu_\mu$), we require that
both muons are positively identified when we look for \cpoddhiggs
candidates in the range
$\higgsmass<1.05$\,\gev. Finally, when selecting candidate events in
the $\eta_b$ mass region $m_{\mu\mu}\sim 9.39$\,\gev, we require that
no secondary photon above a CM energy of $E_2^*=0.08$\,\gev is present
in the event; this requirement suppresses decay chains
$\Y3S\to\gamma_2\chi_b(2S)\to\gamma_1\gamma_2\Y1S$, in which the photon
$\gamma_2$ has a typical CM energy of 100 MeV. 

We use Monte Carlo 
samples generated at 20 values of \higgsmass over a broad range
$0.212<\higgsmass\le9.5$\,GeV of 
possible $A^0$ masses to measure selection efficiency for the signal
events. The efficiency varies between 24-44\%, depending on the
dimuon invariant mass. 

\section{EXTRACTION OF SIGNAL YIELDS}

\begin{figure}[tb]
\begin{center}
\epsfig{file=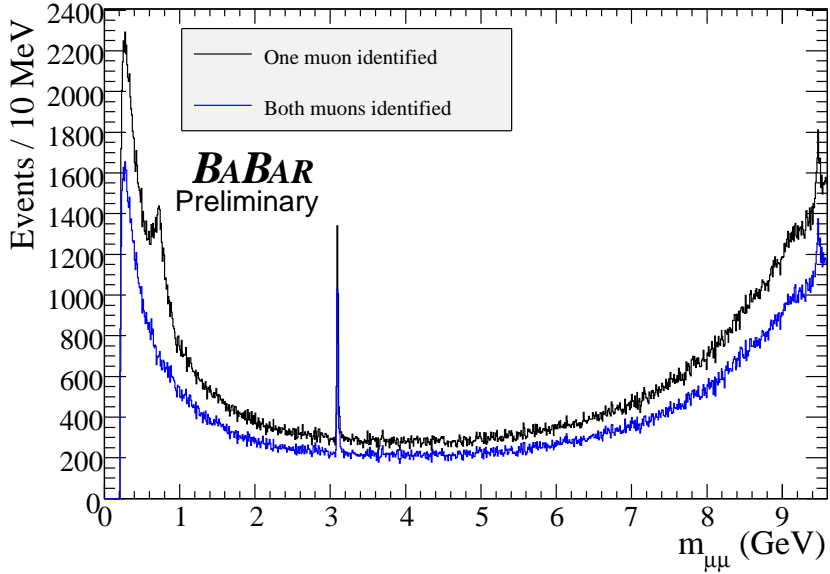, height=3.2in}
\end{center}
\caption{Distribution of the dimuon invariant mass $m_{\mu^+\mu^-}$ in
  the \Y3S data. Black histogram shows the distribution for the
  selection in which only one of two muons is required to be
  positively identified. The peak from
  $e^+e^-\to\gamma_\mathrm{ISR}\rho^0(770)$, $\rho^0\to\pi^+\pi^-$, in
  which one of the pions is misidentified as a muon, is clearly
  visible. 
  Blue (lower) histogram shows the distribution for the
  selection in which both muons are
  positively identified. The ISR-produced peaks at $J/\psi$ and $\Y1S$
  masses are visible. 
}
\label{fig:spectrum_run7}
\end{figure}

The invariant mass spectrum for the selected candidates in the \Y3S
dataset is shown in Fig.~\ref{fig:spectrum_run7}. 
We extract the yield of signal events as a function of the assumed
mass $m_\cpoddhiggs$ in the interval $0.212\le m_\cpoddhiggs\le 9.3$\,GeV by
performing a series of unbinned extended maximum likelihood fits
to the distribution of the ``reduced mass''
\beq
\redmass = \sqrt{m_{\mu\mu}^2 - 4m_\mu^2}\ . 
\eeq
The choice of this variable is motivated by the distribution of the 
{\em continuum background\/} from $e^+e^-\to\gamma\mu^+\mu^-$, which
is a smooth 
function of \redmass across the entire range of
interest, in particular, the region near the
kinematic threshold $m_{\mu\mu}\approx 2m_\mu$ ($\redmass\approx0$). 
Each fit is performed over a small range of \redmass around the value
expected for a particular \higgsmass. 
We use the  \Y4S
sample to determine the probability density functions (PDFs) for the
continuum background in each fit window, which agree within statistical
uncertainties with Monte 
Carlo simulations. We use a threshold (hyperbolic) function to
describe the background below $\redmass<0.23$\,\gev; its parameters are
fixed to the values determined from the fits to the  \Y4S dataset. 
Elsewhere the
background is well described in each limited \redmass range by a first-order
($\redmass<9.3$\,GeV) or second-order ($\redmass>9.3$\,GeV) polynomial. 

The signal PDF is described by a sum of two Crystal Ball functions~\cite{ref:CBshape}
 with tail parameters on either side of the maximum. The
signal PDFs are centered around the expected values of
$\redmass=\sqrt{m^2_{\cpoddhiggs}- 4m_\mu^2}$ and have the typical
resolution of $2-10$\,MeV, which increases monotonically with \higgsmass. 
We determine the PDF as a function of
$\higgsmass$ using a set of high-statistics simulated samples of signal
events, and we interpolate PDF parameters and signal efficiency values
linearly between simulated
points. We determine the uncertainty in the PDF parameters by
comparing the distributions of the simulated and reconstructed 
$e^+e^-\to\gamma_\mathrm{ISR}J/\psi$, $J/\psi\to\mu^+\mu^-$
events. 

Known resonances, such as \jpsi, $\psi(2S)$, and \Y1S, are present in our
sample in specific intervals of $\redmass$, and constitute {\em peaking
  background\/}.  
We include these contributions in the fit where appropriate, and describe the
shape of the resonances using the same functional form as for the signal,
a sum of two 
Crystal Ball functions, with parameters determined from the dedicated
MC samples. We do not search for \cpoddhiggs signal in the immediate
vicinity of \jpsi and $\psi(2S)$, ignoring the region of $\approx\pm40$\,MeV
around \jpsi (approximately $\pm5\sigma$) and $\approx\pm25$\,MeV
($\approx\pm3\sigma$) around $\psi(2S)$. 

For each assumed value of \higgsmass, we perform a likelihood fit to
the $\redmass$ distribution under the following conditions:
\begin{itemize}
\item $0.212\le\higgsmass<0.5$\,GeV: we use a fixed interval
  $0.01<\redmass<0.55$\,GeV. The fits are done in 2 MeV steps in 
  $\higgsmass$. We use a threshold function to describe the
  combinatorial background PDF below 
  $\redmass<0.23$\,\gev, and constrain it to the shape determined from
  the large \Y4S dataset. For $\redmass>0.23$\,\gev, we describe the
  background by a first-order Chebyshev polynomial and float its
  shape, while requiring continuity at $\redmass=0.23$\,\gev. Signal
  and background yields are free parameters in the fit. 
\item $0.5\le\higgsmass<1.05$\,GeV: we use sliding intervals
  $\mu-0.2<\redmass<\mu+0.1$\,GeV (where $\mu$ is the mean of the signal
  distribution of $\redmass$). We perform fits in 3 MeV
  steps in \higgsmass. First-order polynomial
  coefficient of the background PDF, signal 
  and background yields are free parameters in the fit.
\item  $1.05\le\higgsmass<2.9$\,GeV: we use sliding intervals
  $\mu-0.2<\redmass<\mu+0.1$\,GeV and perform fits in 5 MeV steps in
  $\higgsmass$. First-order polynomial
  coefficient of the background PDF, signal
  and background yields are free parameters in the fit.
\item  $2.9\le\higgsmass\le3.055$\,GeV and 
$3.135\le\higgsmass\le3.395$\,GeV: we use a fixed interval 
$2.7<\redmass<3.5$\,GeV; 5 MeV steps in
  $\higgsmass$. First-order polynomial
  coefficient of the background PDF, signal, $J/\psi$, and background 
  yields  are free parameters in the fit.
\item  $3.4\le\higgsmass<3.55$\,GeV: we use sliding intervals
  $\mu-0.2<\redmass<\mu+0.1$\,GeV and perform fits in 5 MeV steps in
  $\higgsmass$. First-order polynomial
  coefficient of the background PDF, signal
  and background yields are free parameters in the fit.
\item  $3.55\le\higgsmass\le3.66$\,GeV and 
$3.71\le\higgsmass<4.0$\,GeV: we use fixed interval 
$3.35<\redmass<4.1$\,GeV; 5 MeV steps in
  $\higgsmass$. First-order polynomial
  coefficient of the background PDF, signal, $\psi(2S)$, and background
  yields  are free parameters in the fit.
\item  $4.0\le\higgsmass<9.3$\,GeV: we use sliding intervals
  $\mu-0.2<\redmass<\mu+0.1$\,GeV; 5 MeV steps in
  $\higgsmass$.  First-order polynomial
  coefficient of the background PDF, signal
  and background yields are free parameters in the fit.
\item $\eta_b$ region ($m_{\eta_b}=9.390$\,GeV): we use a fixed interval 
$9.2<\redmass<9.6$\,GeV. We constrain the contribution from
  $e^+e^-\to\gamma_\mathrm{ISR}\Y1S$ to the expectation from the \Y4S
  dataset ($436\pm50$ events). Background PDF shape (second-order
  Chebyshev polynomial), 
  yields of $\Y3S\to\gamma\chi_b(2P)\to\gamma\gamma\Y1S$, signal
  $\Y3S\to\gamma\eta_b$ events, and background yields are free
  parameters in the fit. 
\end{itemize}
The step sizes in each interval correspond approximately to the
resolution in $\higgsmass$. 

As a crosscheck, we also perform a set of two-dimensional maximum
likelihood fits to the joint distribution of $\redmass$ and
$\cos\theta^{*}_{\mu\mu}$, where $\theta^{*}_{\mu\mu}$ is the CM
polar angle of the dimuon pair. For scalar \cpoddhiggs, the angular
distribution of the signal events is expected to be
$1+\cos^2\theta^{*}_{\mu\mu}$, modulo acceptance and efficiency
effects. The distribution of the most dominant QED backgrounds is more
strongly peaked in the forward and backward directions. Thus, the
angular distribution can help distinguish any signal peaks from the QED
background. We find that the results of the two-dimensional fit are
consistent with the one-dimensional fit to \redmass\ only. 


\section{SYSTEMATIC UNCERTAINTIES}
\label{sec:Systematics}

The largest systematic uncertainty in $\BR(\Y3S\to\gamma\cpoddhiggs)$
comes from the 
measurement of the selection efficiency. We compare the overall
selection efficiency between the data and the Monte Carlo simulation by
measuring the absolute cross section $d\sigma/d\redmass$ for the
radiative QED process
$e^+e^-\to\gamma\mu^+\mu^-$ over the broad kinematic range
$0<\redmass\le9.6$\,GeV, using a sample of $2.4\,\mathrm{fb}^{-1}$
collected 30\,MeV below the \Y3S. We use the ratio of measured to
expected cross 
sections to correct the signal selection efficiency as a function of
\higgsmass. This correction reaches up to 20\% at low values of
\higgsmass. We use half of the applied correction, or its statistical
uncertainty of 2\%, whichever is larger, as the systematic uncertainty
on the signal efficiency. 
This uncertainty accounts for
effects of selection efficiency, 
reconstruction efficiency (for both charged tracks and the photon),
trigger efficiency, and the uncertainty in estimating the integrated
luminosity. We find the largest difference between the data and
Monte Carlo simulation in modeling of muon identification efficiency. 

We determine the uncertainty in the signal and peaking background PDFs
by comparing the data 
and simulated distributions of $e^+e^-\to\gamma_\mathrm{ISR}J/\psi$
events. We correct for the observed 24\% difference ($5.3$\,MeV in the
simulations versus $6.6$\,MeV in the data) in the width of
the \redmass distribution for these events, 
and use half of the correction to estimate 
the systematic uncertainty on the signal yield. This is the dominant
systematic uncertainty 
on the signal yield for $\higgsmass>0.4$\,\gev. 
Likewise, we find that changes in
the tail parameters of the Crystal Ball PDF describing the $J/\psi$ peak
lead to variations in event yield of less than 1\%. We use this
estimate as a systematic error in the signal yield due to
uncertainty in tail parameters. 

We find excellent agreement in the shape of the continuum background
distributions for $\redmass<0.23$\,\gev between \Y3S and \Y4S
data. We determine the PDF in the fits to \Y4S data, and propagate
their uncertainties to the \Y3S data, where these contributions do not
exceed $\sigma(\BR)=0.3\times10^{-6}$. For the higher masses 
$\redmass>0.23$\,\gev, the background PDF parameters are floated in the
likelihood fit. 

We test for possible bias in the fitted value of the signal yield with a
large ensemble of pseudo-experiments. For each experiment, we generate
a sample of background events according to the number and the PDF
observed in the data, and add a pre-determined number of signal events
from fully-reconstructed signal Monte Carlo samples. The bias is
consistent with zero for all values of \higgsmass, and we assign a
branching fraction 
uncertainty of $\sigma(\BR)=0.02\times10^{-6}$ at all values of
\higgsmass to cover the
statistical variations in the results of the test. 

\begin{figure}[tb]
\begin{center}
\epsfig{file=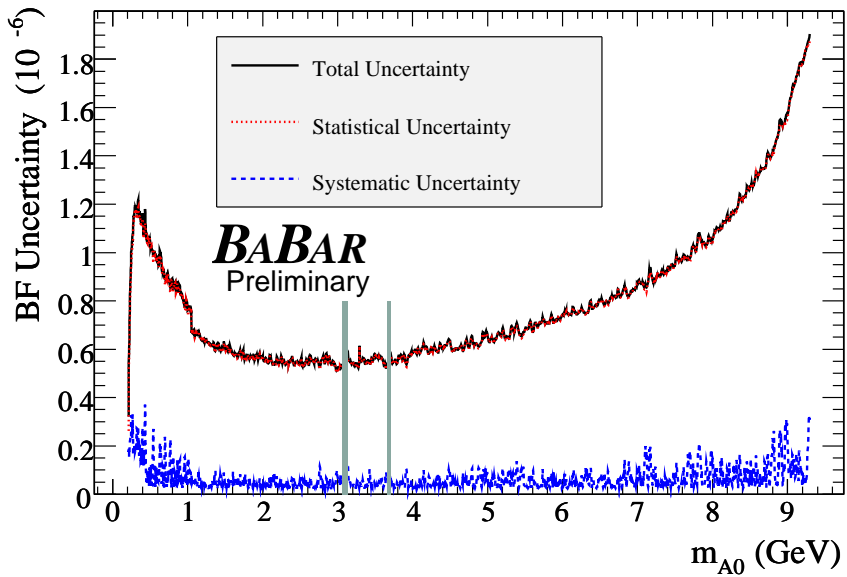, height=3.2in}
\end{center}
\caption{Statistical and systematic uncertainty on the product of branching
  fractions
  $\mathcal{B}(\Y3S\to\gamma\cpoddhiggs)\times\mathcal{B}(\cpoddhiggs\to\mu^+\mu^-)$
  as a function of \higgsmass, extracted from the fits to the \Y3S
  data. Statistical errors are shown as red dot-dashed line, 
  systematic uncertainties are shown as blue dotted line, and the
  total uncertainty, computed as a quadrature sum of statistical and
  systematic errors, is the solid black line. The shaded areas show the regions
  around the \jpsi and $\psi(2S)$ resonances excluded from
  the search. 
}
\label{fig:scan1d_syst_Run7}
\end{figure}

The uncertainties in PDF parameters of both signal and background
and the bias uncertainty affect the signal yield (and therefore
significance of any peak); 
signal efficiency uncertainty does not. The effect of the systematic
uncertainties on the signal yield is 
generally small. The statistical and systematic
uncertainties on the branching fraction
$\BR(\Y3S\to\gamma\cpoddhiggs)$ as a function of \higgsmass are shown
in Fig.~\ref{fig:scan1d_syst_Run7}. 

\section{STATISTICAL INTERPRETATION}
\label{sec:significance}

\begin{figure}[tb]
\begin{center}
\epsfig{file=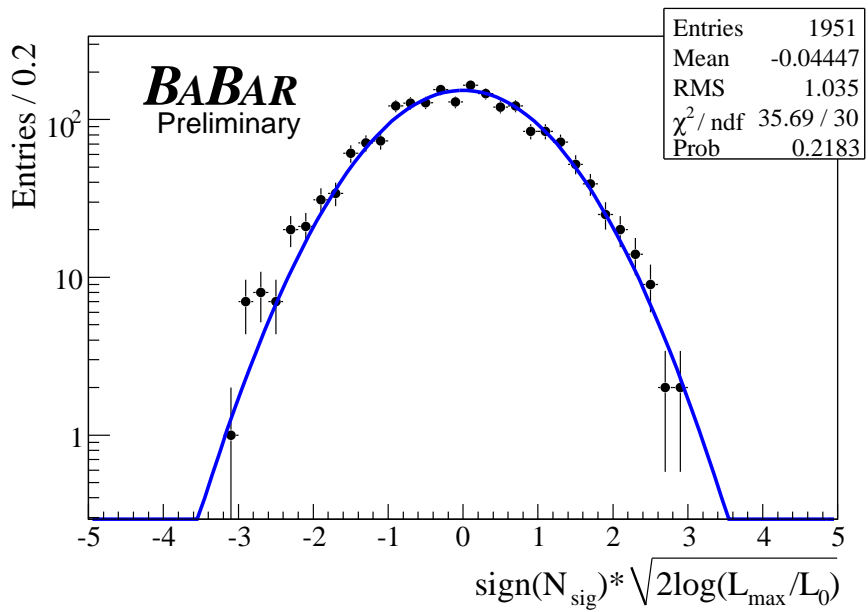, height=3.2in}
\end{center}
\caption{Distribution of the likelihood ratio variable
  $\mathcal{S}$ (with additive systematic uncertainties included for
  the fits to  the \Y3S dataset. The blue curve is the
  Gaussian fit  with fixed $\mu=0$ and $\sigma=1$.  
}
\label{fig:scan1d_sign_Run7}
\end{figure}

In the event of a positive observation, statistical significance of a
particular peak needs to be assessed. Conventionally, this is done by
computing the likelihood ratio variable 
\begin{equation}
\mathcal{S}(\higgsmass) = \sqrt{2\log(L_{\max}/L_0)}
\label{eq:naiveSign}
\end{equation}
where $L_{\max}$ is the maximum
likelihood value for a fit with a floated signal yield centered at
\higgsmass, and $L_0$ is the value of the likelihood for the fixed
zero signal yield. 
Under the null hypothesis (no signal events in the data), the
signed quantity, $sign(N_\mathrm{sig})\mathcal{S}$
is expected to be normal-distributed (where $N_\mathrm{sig}$ is the
fitted signal yield). The distribution for our \Y3S
dataset is shown in Fig.~\ref{fig:scan1d_sign_Run7}. 
Since our scans have $\mathcal{O}(2000)$ \higgsmass
points, we should expect several statistical fluctuations at
the level of $\mathcal{S}\approx 3$, even for a null signal
hypothesis. 

For a single experiment consisting of
$N_\mathrm{trial}$ uncorrelated measurements, the probability to
observe $\mathcal{S}\ge\mathcal{S}_{\max}$ and $N_\mathrm{sig}>0$ is 
\begin{equation}
P(\mathcal{S}_{\max};N_\mathrm{trial}) \approx N_\mathrm{trial}P(\mathcal{S}_{\max};1)
= N_\mathrm{trial}\frac{\mathrm{Erfc}(\mathcal{S}_{\max}/\sqrt{2})}{2}
\label{eq:scanProb}
\end{equation}
(this approximation is good for
$P(\mathcal{S}_{\max};N_\mathrm{trial}) \ll 1$). 
The ``trial factor'' $N_\mathrm{trial}$ is difficult to estimate
analytically. Instead, we determine $N_\mathrm{trial}$ by inspecting
results of $10^8$ Monte Carlo pseudo-experiments;
for each experiment we generate 1951 random values $x_i$
according to 
\begin{equation}
x_i = x_{i-1}\rho + r \sqrt{1-\rho^2}
\label{eq:scanCorr}
\end{equation}
where $\rho=0.84$ is the average correlation coefficient between adjacent
``bins'' $x_i$ and $x_{i-1}$, as determined by \Y3S fits, and $r$
is a normal-distributed random number. We then compute the
maximum value $\mathcal{S}_{\max}$ for each pseudo-experiment. The
cumulative distribution of $\mathcal{S}_{\max}$ describes the chance that a 
pure background would fluctuate to produce a signal peak {\em at any
  value\/} of \higgsmass with 
the likelihood ratio variable $\mathcal{S}\ge\mathcal{S}_{\max}$. 
From the fit to the distribution of $\mathcal{S}_{\max}$ from $10^8$
pseudo-experiments, we determine the effective trial factor
$N_\mathrm{trial}=1436\pm16$. Finally, we convert the likelihood
ratio variable $\mathcal{S}$ to the true
statistical significance $\mathcal{S}^{'}$ in terms of gaussian $\sigma$:
\begin{equation}
\mathcal{S}^{'} \approx
\sqrt{2}\mathrm{Erfc}^{-1}\left[N_\mathrm{trial}\mathrm{Erfc}(\mathcal{S}_{\max}/\sqrt{2})\right]\ (\mathcal{S}>4)
\end{equation}
In 
particular, the threshold for observing ``evidence'' for signal at
{\em any\/} value of \higgsmass is $\mathcal{S}^{'}\ge3.0$; this corresponds
to the likelihood ratio at a {\em particular\/} value of \higgsmass of
$\mathcal{S}\ge4.8$.

\section{RESULTS AND CONCLUSIONS}
\label{sec:Results}

\begin{figure}[b!]
\begin{center}
\epsfig{file=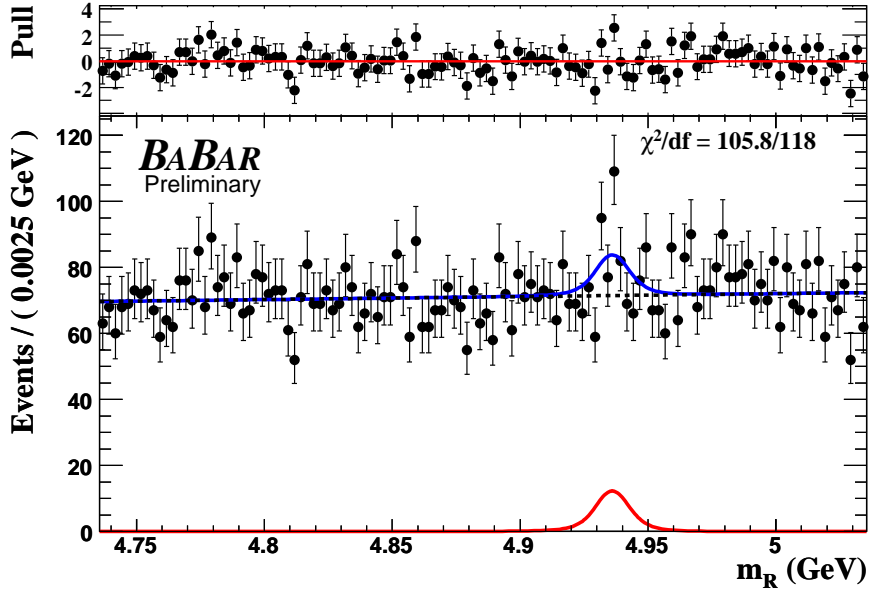, height=3.2in}
\end{center}
\caption{The fit for $\higgsmass=4.940$\,\gev in
  \Y3S dataset. The bottom graph shows the $\redmass$ distribution
  (solid points), overlaid by the full PDF (solid blue
  line). Also shown are the contributions from the signal at
  $\higgsmass=4.940$\,\gev (solid red line) and the continuum
  background (dashed black line). The top plot shows the normalized residuals
  $p=(\mathrm{data}-\mathrm{fit})/\sigma(\mathrm{data})$  with unit
  error bars. The signal peak corresponds to the 
  likelihood ratio variable $\mathcal{S}=3.0$, including
  systematics, and $\BR=(1.9\pm0.7\pm0.1)\times10^{-6}$. 
}
\label{fig:proj1d_4.940}
\end{figure}

\begin{figure}[t!]
\begin{center}
\epsfig{file=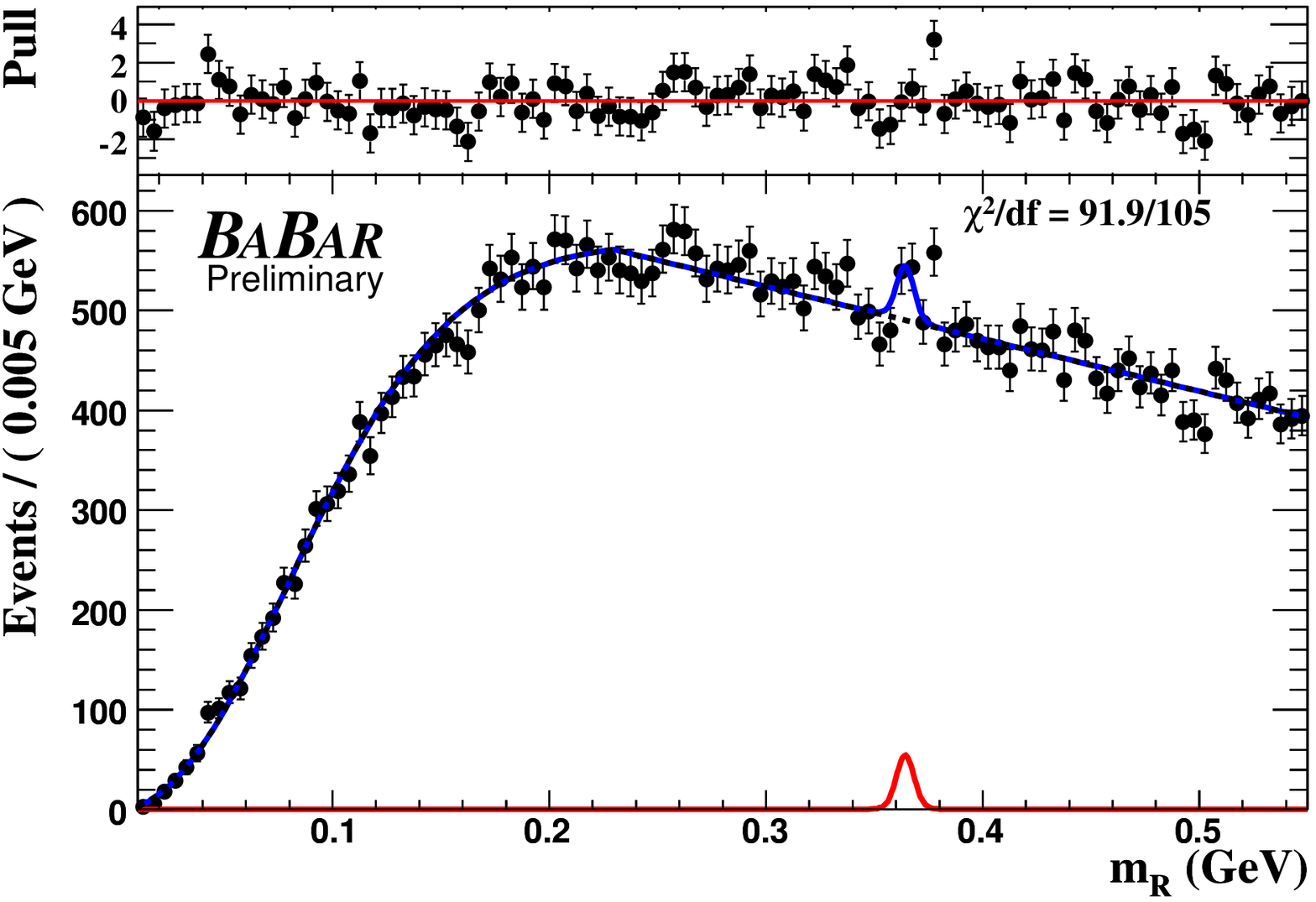, height=3.2in}
\end{center}
\caption{The fit for $\higgsmass=0.426$\,\gev in
  \Y3S dataset. The bottom graph shows the $\redmass$ distribution
  (solid points), overlaid by the full PDF (solid blue
  line). Also shown are the contributions from the signal at
  $\higgsmass=0.426$\,\gev (solid red line) and the continuum
  background (dashed black line). The top plot shows the normalized residuals
  $p=(\mathrm{data}-\mathrm{fit})/\sigma(\mathrm{data})$  with unit
  error bars. The signal peak corresponds to the 
  likelihood ratio variable $\mathcal{S}=2.9$, including
  systematics, and $\BR=(3.1\pm1.1\pm0.3)\times10^{-6}$. 
}
\label{fig:proj1d_0.426}
\end{figure}

For a small number of fits in the scan over the 
\Y3S dataset, we observe local likelihood ratio values $\mathcal{S}$ of
about $3\sigma$. 
The most significant peak is at $\higgsmass=4.940\pm0.003$\,\gev 
(likelihood ratio value $\mathcal{S}=3.0$, including
systematics; $\BR=(1.9\pm0.7\pm0.1)\times10^{-6}$). The second
most-significant peak is at
$\higgsmass=0.426\pm0.001$\,\gev (likelihood ratio 
value $\mathcal{S}=2.9$, including systematics;
$\BR=(3.1\pm1.1\pm0.3)\times10^{-6}$). 
The 
plots for these points are shown in Fig.~\ref{fig:proj1d_4.940} and 
Fig.~\ref{fig:proj1d_0.426}. 
The peak at $\higgsmass=4.940$\,\gev is
theoretically disfavored (since it is significantly above the $\tau$
threshold), while the peak at $\higgsmass=0.426$\,\gev is in the
range predicted by the axion model~\cite{NomuraThaler}. Neither of the
peaks are significant, however, when we take into account the trial
factor discussed in Section~\ref{sec:significance}. At least 80\% of
our pseudo-experiments contain a fluctuation with $\mathcal{S}=3\sigma$
or more.  

\begin{figure}[tb]
\begin{center}
\epsfig{file=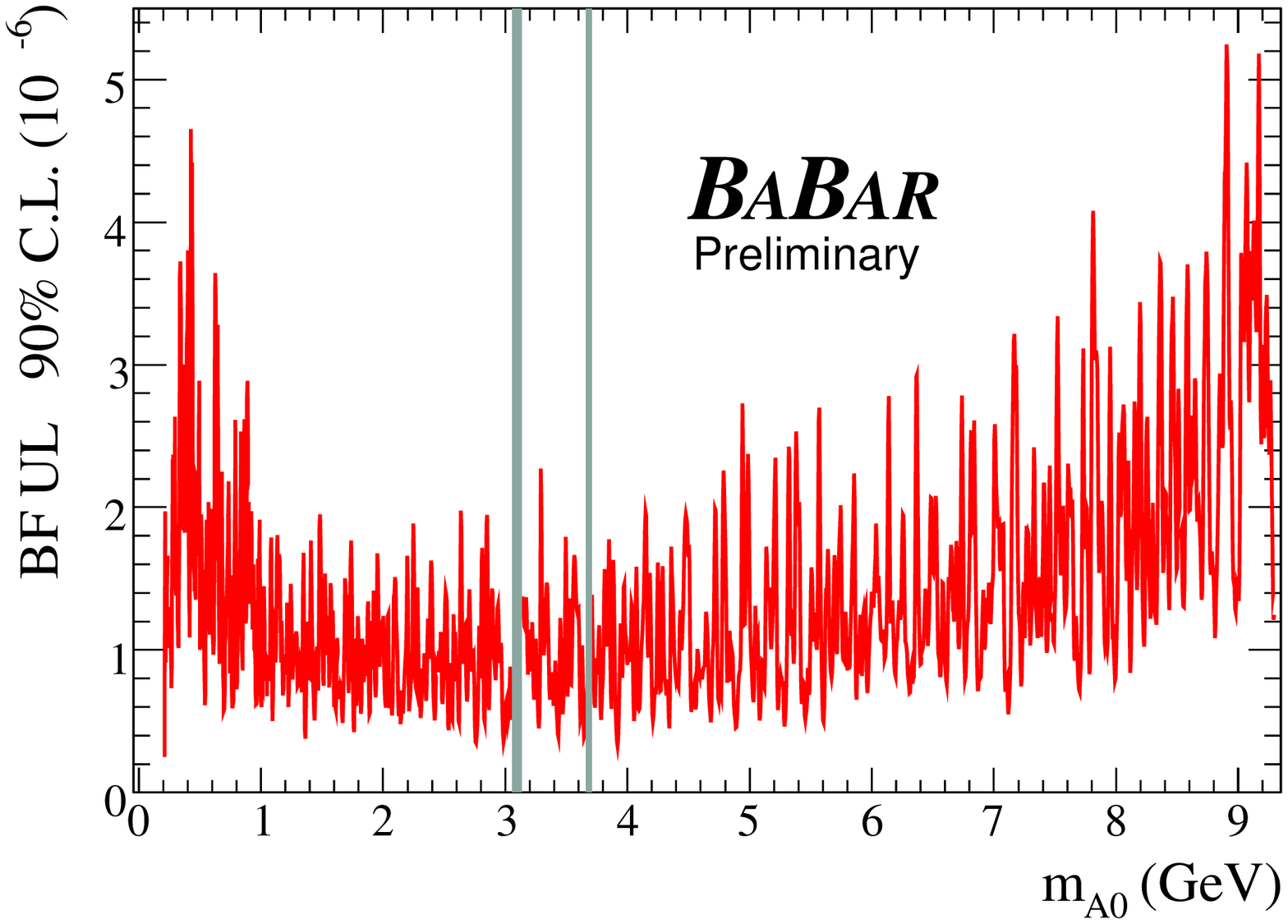, height=3.4in}
\end{center}
\caption{Upper limits on the product of branching fractions $\BR(\Y3S\to\gamma
  A^0)\times\BR(A^0\to\mu^+\mu^-)$ as a function of \higgsmass from
  the fits to \Y3S data. The shaded areas show the regions
  around the \jpsi and $\psi(2S)$ resonances excluded from
  the search. 
}
\label{fig:scan1d_limits}
\end{figure}

\begin{figure}[b!]
\begin{center}
\epsfig{file=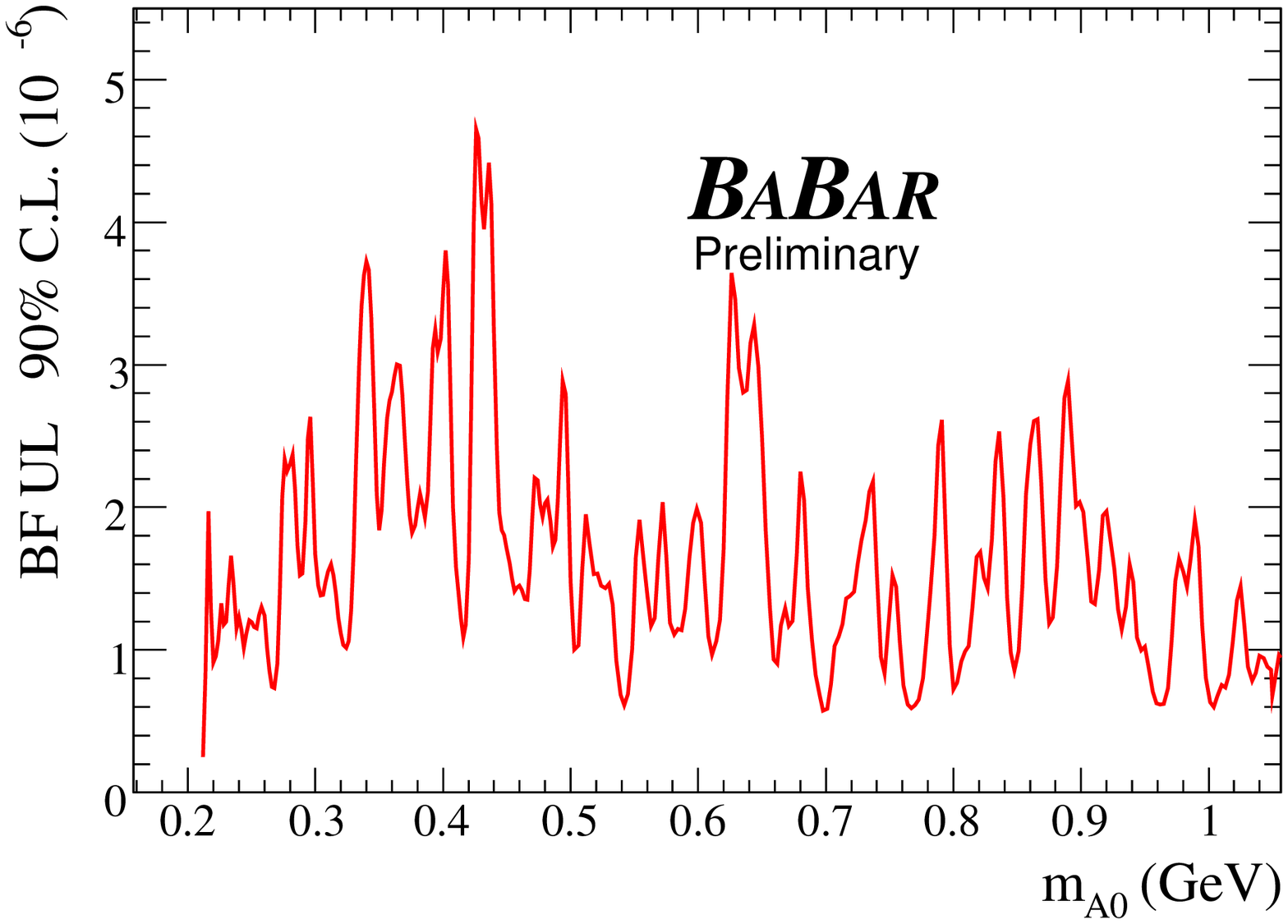, height=3.4in}
\end{center}
\caption{Upper limits on the product of branching fractions $\BR(\Y3S\to\gamma
  A^0)\times\BR(A^0\to\mu^+\mu^-)$ as a function of \higgsmass  in the range
  $0.212\le\higgsmass\le1.05$\,\gev from
  the fits to \Y3S data. 
}
\label{fig:scan1d_limits_low}
\end{figure}

\begin{figure}[tb]
\begin{center}
\epsfig{file=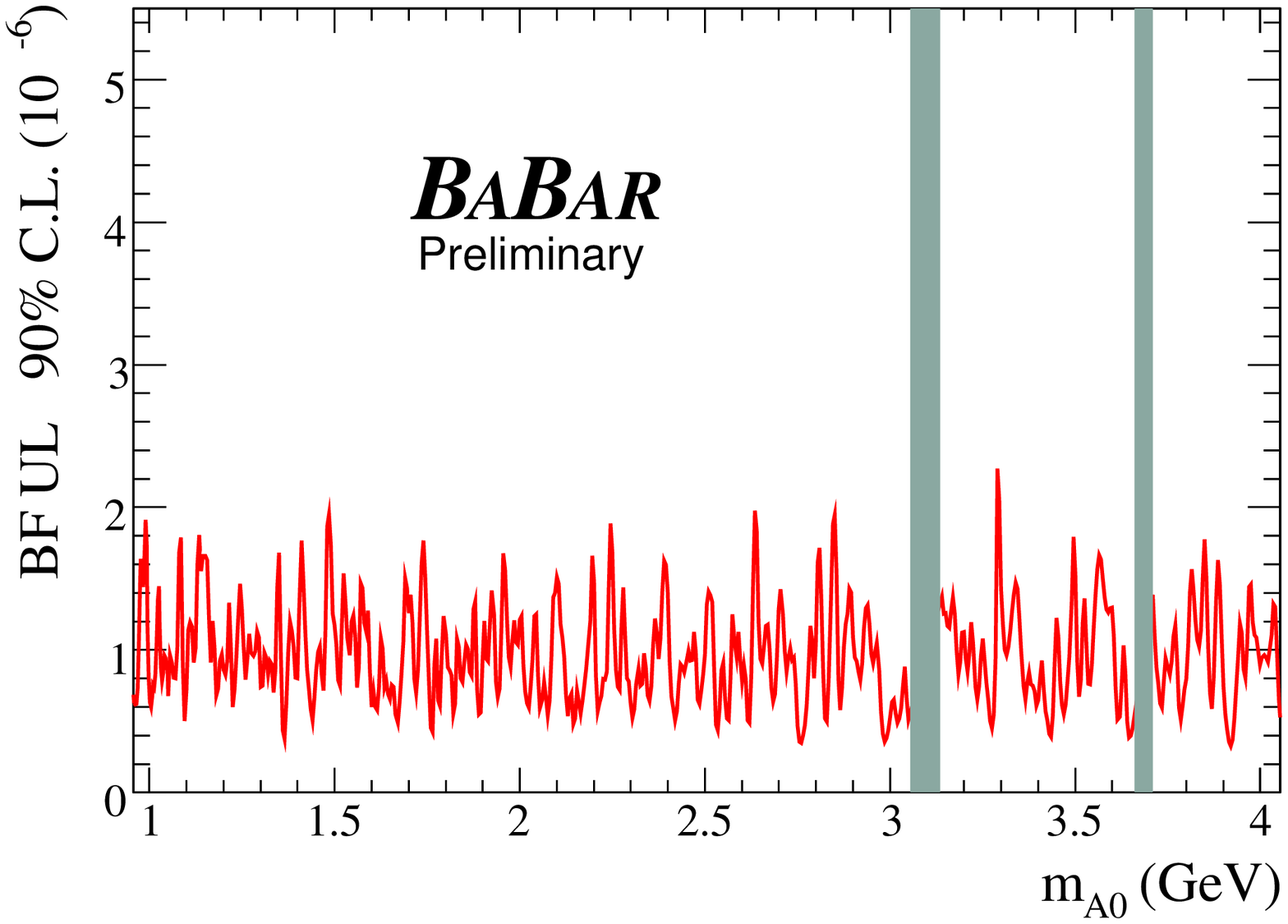, height=3.4in}
\end{center}
\caption{Upper limits on the product of branching fractions $\BR(\Y3S\to\gamma
  A^0)\times\BR(A^0\to\mu^+\mu^-)$ as a function of \higgsmass   in the range
  $1\le\higgsmass\le4$\,\gev from
  the fits to \Y3S data.  The shaded areas show the regions
  around the \jpsi and $\psi(2S)$ resonances excluded from
  the search. 
}
\label{fig:scan1d_limits_med}
\end{figure}

\begin{figure}[tb]
\begin{center}
\epsfig{file=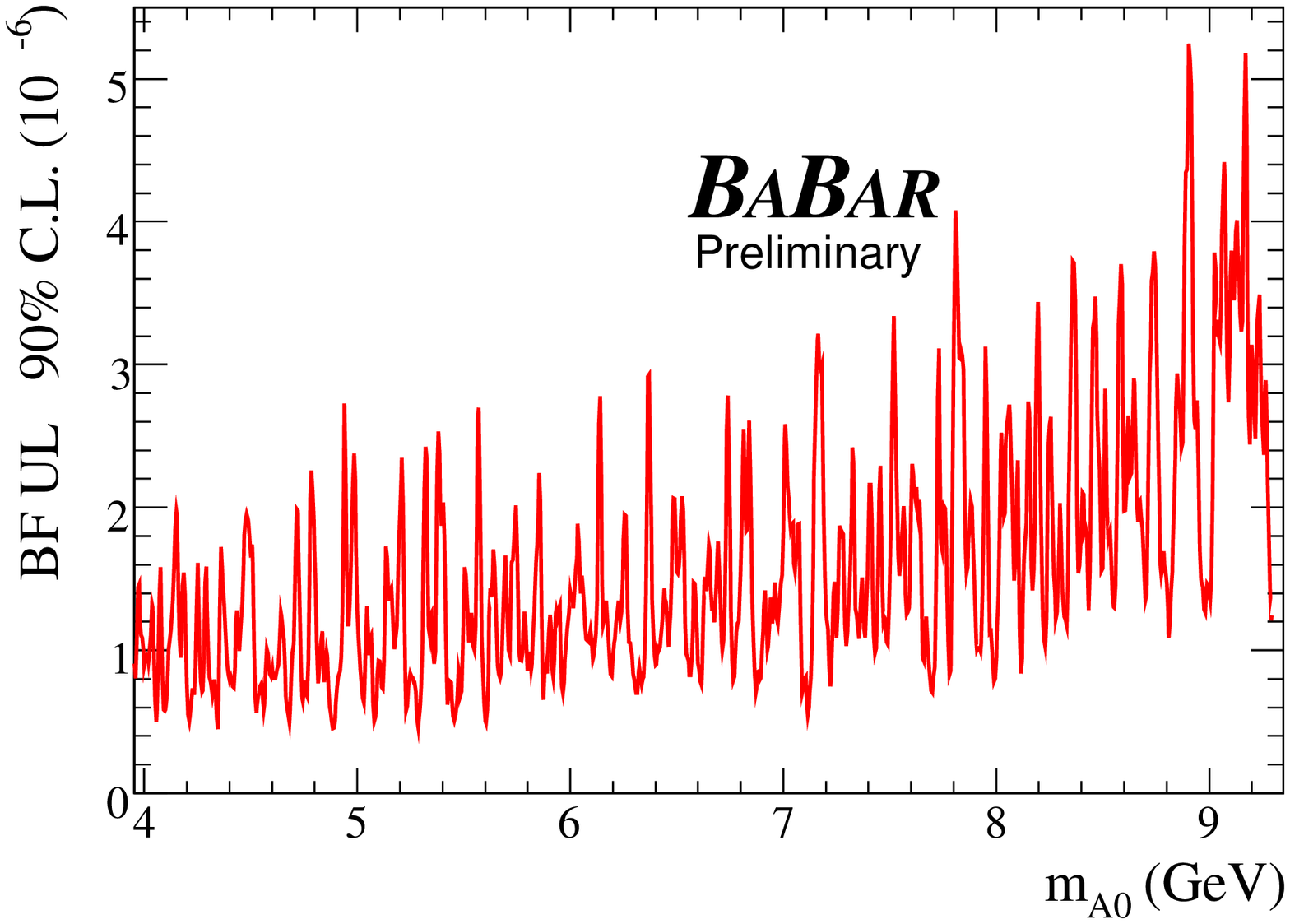, height=3.4in}
\end{center}
\caption{Upper limits on the product of branching fractions $\BR(\Y3S\to\gamma
  A^0)\times\BR(A^0\to\mu^+\mu^-)$ as a function of \higgsmass  in the range
  $4\le\higgsmass\le 9.3$\,\gev from
  the fits to \Y3S data. 
}
\label{fig:scan1d_limits_high}
\end{figure}

Since we do not observe a significant excess of events above the background
in the range $0.212<\higgsmass\le 9.3$\,GeV, we set 
upper limits on the branching fraction
$\mathcal{B}(\Y3S\to\gamma\cpoddhiggs)\times\mathcal{B}(\cpoddhiggs\to\mu^+\mu^-)$.
We add statistical and systematic uncertainties (which include the
additive errors on the signal yield and multiplicative uncertainties
on the signal efficiency and the number of recorded \Y3S decays) in
quadrature. 
The 90\% C.L. Bayesian upper limits,
computed with a uniform prior and assuming a Gaussian likelihood
function, are shown in Fig.~\ref{fig:scan1d_limits}-\ref{fig:scan1d_limits_high} as a function of mass
$\higgsmass$. The limits fluctuate depending on the central value of
the signal yield returned by a particular fit, and range from
$0.25\times10^{-6}$ to $5.2\times10^{-6}$. 

\begin{figure}[tb]
\begin{center}
\epsfig{file=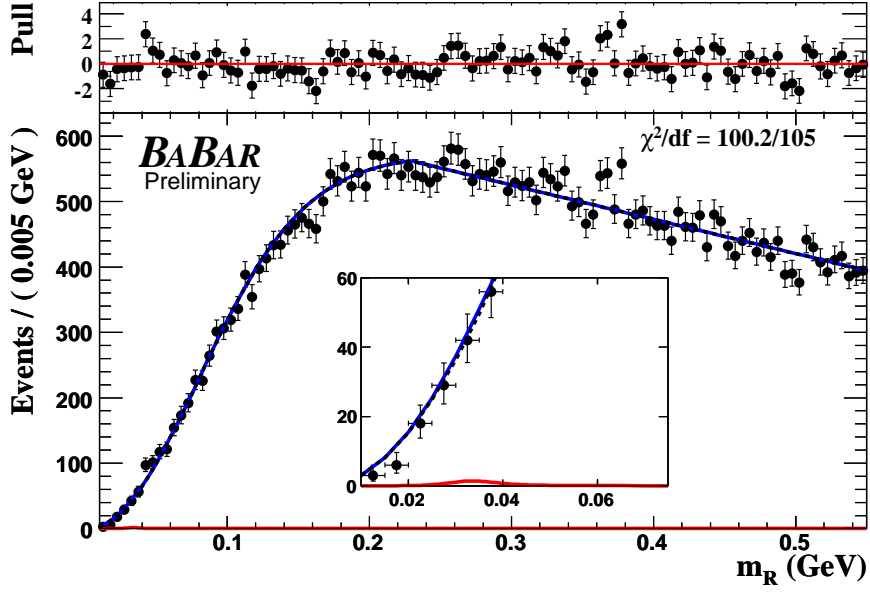, height=3.2in}
\end{center}
\caption{The fit for $\higgsmass=0.214$\,\gev
  (HyperCP candidate) in
  the \Y3S dataset. The bottom graph shows the $\redmass$ distribution
  (solid points), overlaid by the full PDF (solid blue
  line). Also shown are the contributions from the signal at
  $\higgsmass=0.214$\,\gev (solid red line) and the continuum
  background (dashed black line). The inset zooms in on the HyperCP
  region of interest. The top plot shows the normalized residuals
  $p=(\mathrm{data}-\mathrm{fit})/\sigma(\mathrm{data})$  with unit
  error bars.  
}
\label{fig:proj1d_0.214}
\end{figure}

We do not observe any significant signal at the HyperCP mass,
$\higgsmass=0.214$\,\gev (see Fig.~\ref{fig:proj1d_0.214}). We find
$\BR(\Y3S\to\gamma A^0(214))=(0.12^{+0.43}_{-0.41}\pm0.17)\times10^{-6}$, 
and set
an upper limit $\BR(\Y3S\to\gamma A^0(214))<0.8\times10^{-6}$ at 90\% C.L. 

\begin{figure}[t]
\begin{center}
\epsfig{file=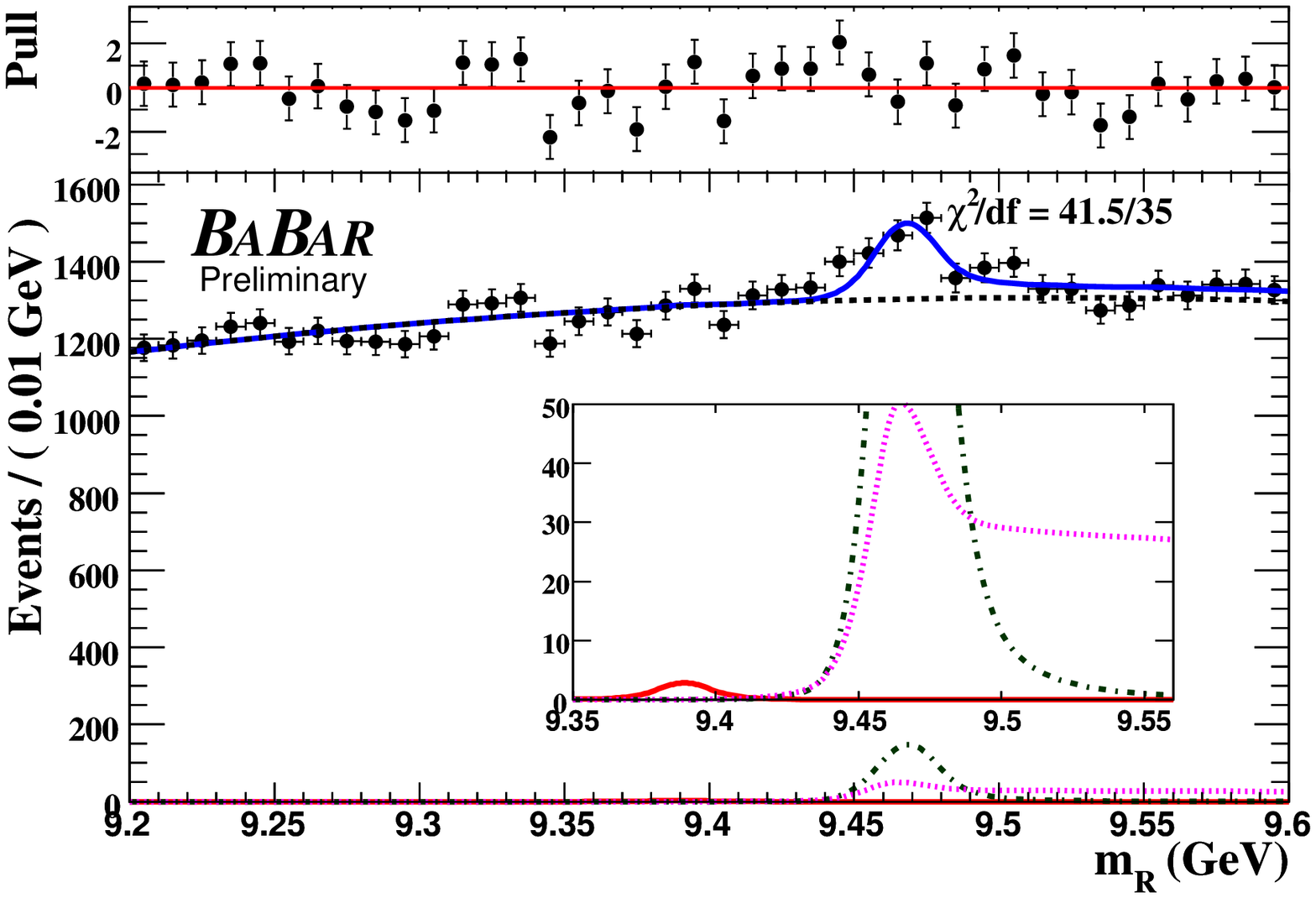, height=3.2in}
\end{center}
\caption{The fit for the $\eta_b$ region in
  \Y3S dataset. The bottom graph shows the $\redmass$ distribution
  (solid points), overlaid by the full PDF (solid blue
  line). Also shown are the contributions from the signal at
  at $m_{\eta_b}=9.389$\,\gev (solid red line), background from the
  $e^+e^-\to\gamma_\mathrm{ISR}\Y1S$ (dot-dashed green line),
  background from $\Y3S\to\gamma\chi_{b}(2P)$,
  $\chi_{b}(2P)\to\gamma\Y1S$ (dotted magenta line), and the continuum
  background (dashed black line). The inset shows the signal, 
  $e^+e^-\to\gamma_\mathrm{ISR}\Y1S$, and $\chi_{b}(2P)\to\gamma\Y1S$
  in more detail. The top plot shows the normalized residuals
  $p=(\mathrm{data}-\mathrm{fit})/\sigma(\mathrm{data})$  with unit
  error bars. 
}
\label{fig:etab_run7}
\end{figure}

A fit to the $\eta_b$ region is shown in Fig.~\ref{fig:etab_run7}. We
find
$\BR(\Y3S\to\gamma\eta_b)\times\BR(\eta_b\to\mu^+\mu^-)=(0.2\pm3.0\pm0.9)\times10^{-6}$,
consistent with zero. Taking into account the \babar\
measurement of $\BR(\Y3S\to\gamma\eta_b)=(4.8\pm0.5\pm1.2)\times10^{-4}$, we
can derive $\BR(\eta_b\to\mu^+\mu^-)=(0.0\pm0.6\pm0.2)\%$, or an upper
limit $\BR(\eta_b\to\mu^+\mu^-)<0.8\%$ at 90\% C.L. This is consistent
with expectations from the quark model. All results above are preliminary. 

The limits we set are more stringent than those reported by the CLEO
collaboration recently~\cite{CLEO2008}. Our limits rule out much of the
parameter space allowed by the light Higgs~\cite{Dermisek:2006py} and
axion~\cite{NomuraThaler} models. 



\clearpage

\section{ACKNOWLEDGMENTS}
\label{sec:Acknowledgments}

We are grateful for the 
extraordinary contributions of our \pep2\ colleagues in
achieving the excellent luminosity and machine conditions
that have made this work possible.
The success of this project also relies critically on the 
expertise and dedication of the computing organizations that 
support \babar.
The collaborating institutions wish to thank 
SLAC for its support and the kind hospitality extended to them. 
This work is supported by the
US Department of Energy
and National Science Foundation, the
Natural Sciences and Engineering Research Council (Canada),
the Commissariat \`a l'Energie Atomique and
Institut National de Physique Nucl\'eaire et de Physique des Particules
(France), the
Bundesministerium f\"ur Bildung und Forschung and
Deutsche Forschungsgemeinschaft
(Germany), the
Istituto Nazionale di Fisica Nucleare (Italy),
the Foundation for Fundamental Research on Matter (The Netherlands),
the Research Council of Norway, the
Ministry of Education and Science of the Russian Federation, 
Ministerio de Educaci\'on y Ciencia (Spain), and the
Science and Technology Facilities Council (United Kingdom).
Individuals have received support from 
the Marie-Curie IEF program (European Union) and
the A. P. Sloan Foundation.

We thank Radovan Dermisek, Jack Gunion, Zoltan Ligeti, Yasunori
Nomura, Miguel Sanchis-Lozano, and Jesse Thaler for stimulating discussions.

\end{document}